\documentclass[12pt,preprint]{aastex}
\usepackage{amssymb}
\usepackage{amsmath}
\usepackage{graphicx}

\defcitealias{egocat}{C08} 	
\defcitealias{maserpap}{C09}

\begin{document}
\shortauthors{Cyganowski et al.}

\title{G11.92$-$0.61-MM2: A Bonafide Massive Prestellar Core?}
\author{C.J. Cyganowski\altaffilmark{1,2}, 
  C.L. Brogan\altaffilmark{3},
  T.R. Hunter\altaffilmark{3}, 
  D. Graninger\altaffilmark{2},
  K.I. {\"O}berg\altaffilmark{2},
  A. Vasyunin\altaffilmark{4,5,6},
  Q. Zhang\altaffilmark{2},
  R. Friesen\altaffilmark{7},
  S. Schnee\altaffilmark{3}
  }

\email{cc243@st-andrews.ac.uk}

\altaffiltext{1}{Scottish Universities Physics Alliance (SUPA), School of Physics and Astronomy, University of St. Andrews, North Haugh, St Andrews, Fife KY16 9SS, UK}
\altaffiltext{2}{Harvard-Smithsonian Center for Astrophysics, Cambridge, MA 02138}
\altaffiltext{3}{NRAO, 520 Edgemont Rd, Charlottesville, VA 22903}
\altaffiltext{4}{School of Physics and Astronomy, University of Leeds, Leeds LS2 9JT, UK}
\altaffiltext{5}{Max-Planck-Institut für Extraterrestrische Physik (MPE), Giessenbachstr.1, D-85748 Garching, Germany}
\altaffiltext{6}{Ural Federal University, Ekaterinburg, Russia}
\altaffiltext{7}{Dunlap Institute for Astronomy and Astrophysics, University of Toronto, 50 St George St., Toronto, ON M5S 3H4, Canada}

\begin{abstract}
Core accretion models of massive star formation require the existence
of stable massive starless cores, but robust observational examples of
such objects have proven elusive.  We report subarcsecond-resolution
SMA 1.3 mm, 1.1 mm, and 0.88 mm and VLA 1.3 cm observations of an excellent
massive starless core candidate, G11.92$-$0.61--MM2, initially
identified in the course of studies of GLIMPSE Extended Green Objects
(EGOs). Separated by $\sim$7\farcs2 from the nearby MM1 protostellar hot core, MM2
is a strong, compact dust continuum source (submillimeter spectral
index $\alpha=$2.6$\pm$0.1), but is devoid of star formation
indicators.  In contrast to MM1, MM2 has no masers, no centimeter
continuum, and no (sub)millimeter wavelength line emission in $\sim$24
GHz of bandwidth observed with the SMA, including N$_{2}$H$^{+}$(3-2),
HCO$^{+}$(3-2), and HCN(3-2).  Additionally, there is no evidence for
an outflow driven by MM2.  The (sub)millimeter spectral energy
distribution (SED) of MM2 is best fit with a dust temperature of $\sim$17-19 K and luminosity of $\sim$5-7 L$_{\odot}$.  The combined
physical properties of MM2, as inferred from its dust continuum
emission, are extreme: M$\gtrsim$30 M$_{\odot}$ within a radius$<$1000
AU, N$_{\rm H_2}>$10$^{25}$ cm$^{-2}$ and n$_{\rm H_2}>$10$^{9}$
cm$^{-3}$.  Comparison of the molecular abundance limits derived from
our SMA observations with gas-grain chemical models indicates that
extremely dense (n(H)$>>$10$^{8}$ cm$^{-3}$), cold ($<$20 K) conditions
are required to explain the lack of observed (sub)millimeter line
emission, consistent with the dust continuum results.  Our data
suggest that G11.92$-$0.61--MM2 is the best candidate for a bonafide
massive prestellar core found to date, and a promising target for
future, higher-sensitivity observations.

\end{abstract}


\section{Introduction}\label{intro}

Do massive starless cores exist in nature?  The answer to this
question is a key discriminant between the two major classes of models
for massive star formation: ``core accretion'' and ``competitive
accretion'' \citep[recently reviewed by][]{Tan14}.  Core accretion models require, as initial conditions,
gravitationally bound, starless massive cores
\citep[e.g.][]{MT02,MT03,Myers13}; competitive accretion
models do not \citep[e.g.][]{Smith09,Bonnell11}.  In the core
accretion scenario, forming a high-mass star (M$_{\rm
  ZAMS}>$8 M$_{\odot}$) requires an initial core mass $\ge$2-3$\times$ larger \citep{Alves07,Rathborne09,Tan14}.

Observationally, examples of massive starless \emph{cores}---$<$0.1 pc structures likely to form single stars or
small multiple systems---have proven elusive.  In addition to the
candidates disqualified by sensitive mid-infrared surveys with
\emph{Spitzer} and \emph{Herschel}, centimeter-submillimeter interferometers have
revealed molecular outflows and/or masers---indisputable signs of
active star formation---in past ``starless'' core candidates
(e.g.\ \citealt{Bontemps10}, \citealt{DuarteCabral13} in Cygnus-X).
In this context, the best
chances for identifying robust massive starless core candidates lie in
massive star-forming regions for which comprehensive, high-resolution
multiwavelength datasets are available.

In studying GLIMPSE Extended Green Objects
\citep[EGOs;][]{egocat}, we have identified an excellent candidate for
a massive starless core: G11.92$-$0.61-MM2.  Our initial Submillimeter
Array (SMA) 1.3 mm observations of the EGO G11.92$-$0.61 revealed a
massive (proto)cluster, containing three compact cores
\citep[][resolution $\sim$2.4\arcsec]{C11}.  MM2 exhibited strong
millimeter continuum emission, but, remarkably, no line emission across
$\sim$4 GHz of SMA bandwidth.  
MM2 also lacks other star formation indicators: MM2 is not
associated with CH$_{3}$OH maser or centimeter continuum emission, does not drive a
molecular outflow, and has no H$_{2}$O maser emission \citep{maserpap,C11,C11vla,HC96,Breen11}.  In this Letter, we present
subarcsecond-resolution SMA\footnote{The Submillimeter Array is a
joint project between the Smithsonian Astrophysical Observatory and
the Academia Sinica Institute of Astronomy and Astrophysics and is
funded by the Smithsonian Institution and the Academia Sinica.}
observations of G11.92$-$0.61 at 1.3, 1.1, and 0.88 mm,
totalling $\sim$24 GHz of bandwidth.  Together with Karl G.\ Jansky Very Large Array (VLA) 1.3 cm NH$_{3}$
and continuum observations, we use these data to constrain the physical and chemical
properties of MM2, and find that MM2 is the best candidate for a
bonafide massive starless core discovered to date.  Throughout, we
adopt the maser parallax distance of 3.37$^{+0.39}_{-0.32}$ kpc \citep{Sato14}.

\section{Observations}\label{obs}

SMA observations of G11.92$-$0.61 were obtained at 1.3, 1.1, and
0.88 mm, as summarized in Table~\ref{obs_table}.  
The data were calibrated and imaged in CASA.  For the 2011 data, system temperature
calibration was first applied in MIRIAD; for the 2013 data, the 
sma2casa filler\footnote{http://www.cfa.harvard.edu/sma/casa/} was
used.  The continuum was estimated in the \emph{uv}-plane, using
line-free channels, and subtracted from the line emission.  The
continuum was then self-calibrated, and the solutions applied to the
line data.  For each dataset, the uniform-spectral-resolution line
data (0.8125 MHz channels) were resampled to a common velocity
resolution (Table~\ref{obs_table}), then Hanning-smoothed.  (To obtain
better spectral resolution on the N$_{2}$H$^{+}$(3-2) line, a
mixed-spectral-resolution mode was employed, see Table~\ref{obs_table}).
The $^{12}$CO(3-2) data were further smoothed to 3 km s$^{-1}$.

The NRAO\footnote{The National Radio Astronomy Observatory is a
  facility of the National Science Foundation operated under
  cooperative agreement by Associated Universities, Inc.} VLA
observations of G11.92$-$0.61 are part of a survey of massive
protostellar objects in 1.3 cm continuum and line emission
\citep[][in prep; here we consider only the
NH$_{3}$ and 1.3 cm continuum data]{rsro,Brogan_IAU}.  The VLA data were calibrated, imaged, and
self-calibrated in CASA.  

Observational parameters and image properties are
listed in Table~\ref{obs_table}.  
All measurements were made from images corrected for the primary beam response.

\section{Results}

\subsection{Continuum Emission}\label{results_cont}

The continuum emission from the two brightest millimeter cores in the G11.92$-$0.61 protocluster, MM1 and
MM2, is detected with high S/N in all of our
new (sub)millimeter images:
observed source properties are summarized in Table~\ref{cont_table}.
Separated from MM2 by only $\sim$7\farcs2 (0.12 pc), MM1 appears to be
a typical hot core \citep[][\S\ref{results_sf}]{C11}, and so
provides a useful basis for comparison.  
For both cores, the $\sim$0\farcs5-resolution Very Extended
configuration (VEX) SMA 1.3 mm image recovers $\sim$40\% of the flux density in the $\sim$2\farcs4-resolution 1.3 mm
SMA image and $\sim$80\% of the flux density in the $\sim$1\farcs1-resolution CARMA 1.4 mm image \citep[both from][]{C11}.
The fitted size of MM1 is consistently smaller than the beam
(Table~\ref{cont_table}), indicating that it is unresolved in all
three (sub)millimeter images.  In contrast, in the highest-resolution SMA
image, the fitted size of MM2 is comparable to the beam,
and the source appears slightly extended (Fig.~\ref{multipanel_fig}).

To measure (sub)millimeter spectral indices, a 1.3 mm
image was made using only those projected baselines spanned by the 1.1
mm data, then convolved to the 1.1 mm synthesized beam.  The 0.88
and 1.1 mm observations have roughly comparable
\emph{uv}-coverage, so the 0.88 mm image
was simply convolved to the 1.1 mm synthesized beam.  Flux densities
measured from these images are presented in Table~\ref{cont_table}.
In fitting spectral indices, we include the statistical uncertainties
(Table~\ref{cont_table}) and conservative estimates of the absolute
flux calibration uncertainty (15\% at 1.3 mm, 20\% at 1.1 and 0.88 mm).  The fitted spectral indices are $\alpha=$3.1$\pm$0.1 for MM1
and $\alpha=$2.6$\pm$0.1 for MM2.

Our 1.3 cm VLA continuum image confirms the presence of CM1
\citep[centimeter-wavelength counterpart to MM1;][]{C11vla} at the
$\sim$8$\sigma$ level (Fig.~\ref{multipanel_fig},
Table~\ref{cont_table}).  Located $\sim$0\farcs07 (240 AU) southwest
of the fitted position of MM1, the (unresolved) 1.3 cm emission is too
strong to be due purely to dust (S$_{\rm dust,1.3cm}\sim$0.2 mJy,
assuming $\alpha=$3.1).
No 1.3 cm counterpart to MM2 is detected: the 4$\sigma$ limit is 0.30
mJy beam$^{-1}$, corresponding to a limiting size r$\lesssim$17 AU for
any optically thick hypercompact HII region
\citep[following][]{C11vla}.

\subsection{(Lack of) Line Emission}\label{results_sf}

The most remarkable characteristic of MM2 is its lack of
(sub)millimeter-wavelength line emission: as shown in Figure~\ref{cont_peak_spectra_fig}, the image cubes 
are devoid of line emission at the MM2 position
across $\sim$24 GHz of bandwidth observed with the SMA.  We searched
the spectra at the MM2 continuum peak 
for $>$4$\sigma$ excursions in
$\ge$2 adjacent channels.  The only feature is at $\nu_{\rm
  observed}=$232.3365 GHz: an unresolved $\sim$5.5$\sigma$ peak
coincident with MM2 ($>$4$\sigma$ in one adjacent channel, $\nu_{\rm
  observed}=$232.3374 GHz).  Searching the splatalogue line catalog
near the expected rest frequency (for V$_{\rm LSR}\sim$35-37 km
s$^{-1}$) returns primarily transitions with E$_{\rm upper}>$100 K,
inconsistent with the lack of other line emission: the only exception
is an unidentified transition, U-232364 ($\nu_{\rm rest}=$232.364 GHz).
Without a plausible identification, the available evidence is
insufficient to conclude that this weak, narrow feature represents
real line emission associated with MM2.
In marked
contrast, MM1 exhibits copious line emission in molecules
characteristic of hot cores \citep[including CH$_{3}$CN, OCS,
  HC$_{3}$N: Fig.~\ref{cont_peak_spectra_fig},][]{C11}.  

\begin{figure}
\centerline{\includegraphics[scale=0.7]{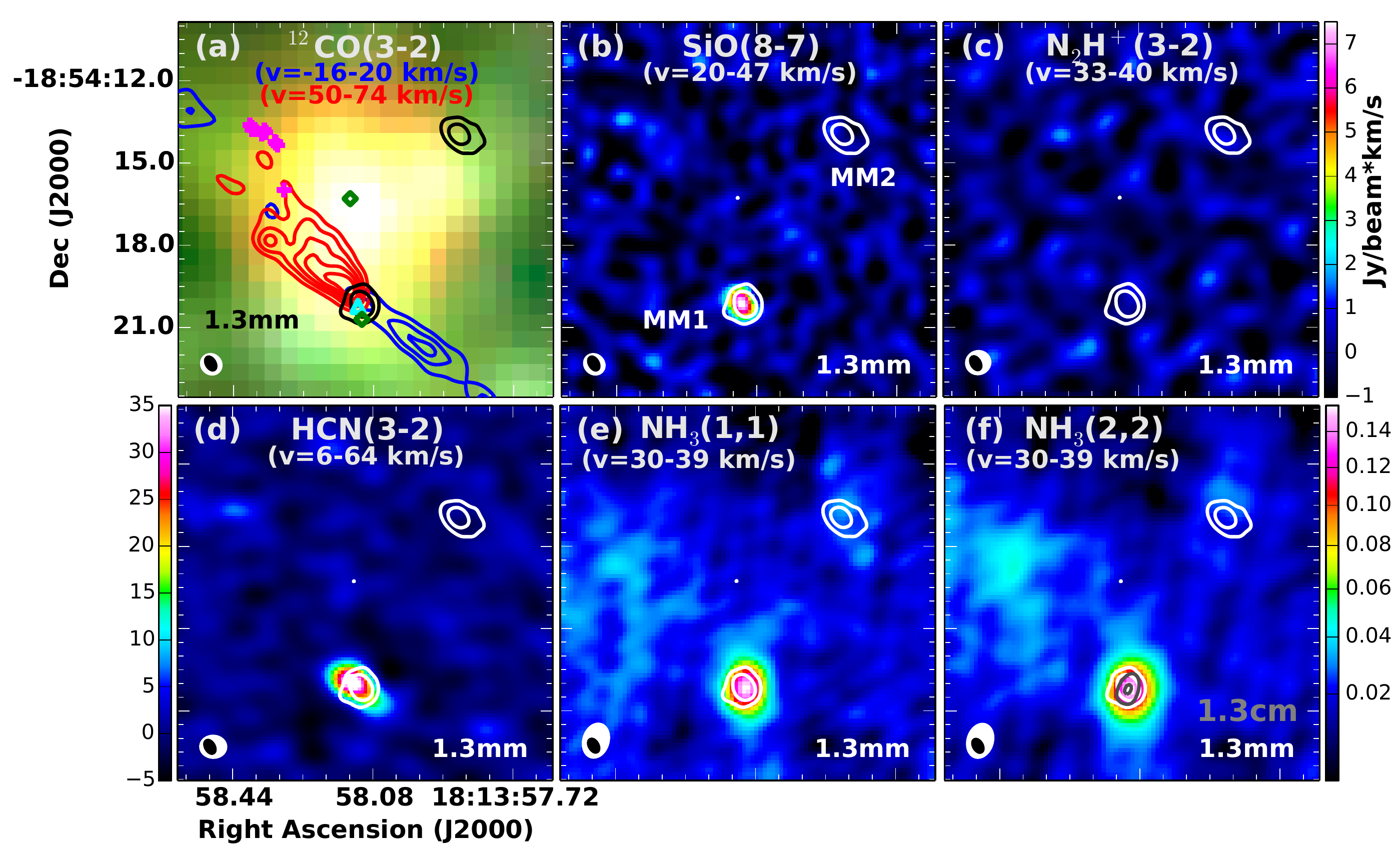}}
\centerline{\includegraphics[scale=1.05]{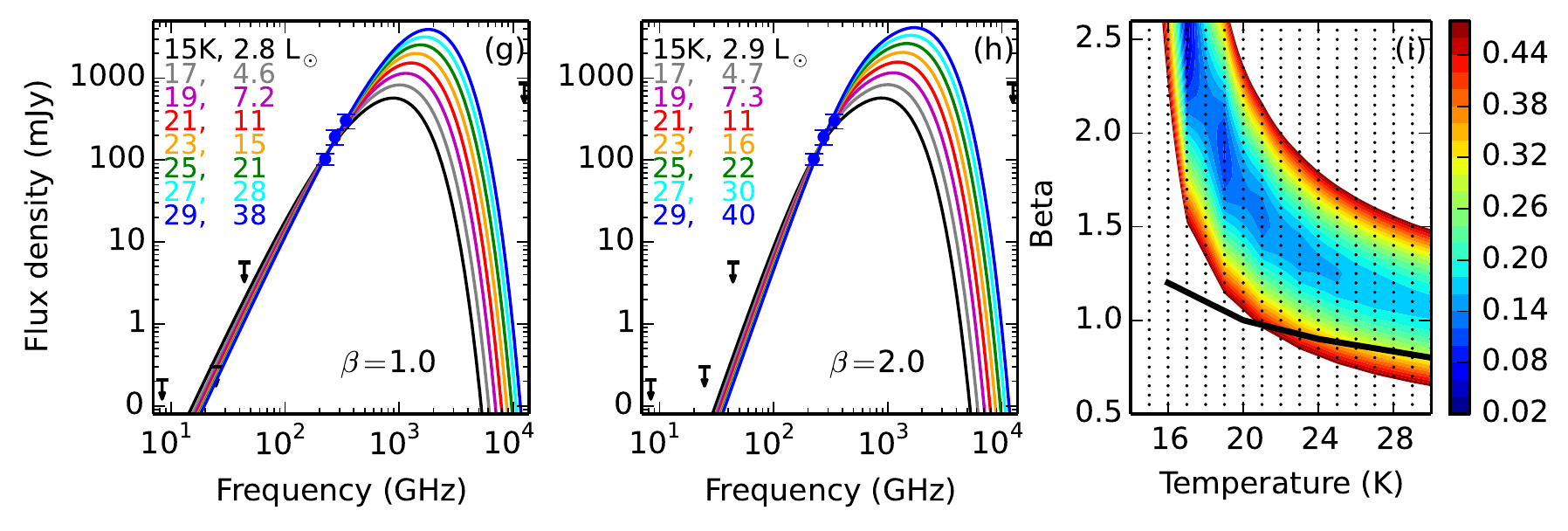}}
\caption{\scriptsize SMA 1.3mm continuum contours
  overlaid on (a) three-color \emph{Spitzer} image (RGB: 8.0,4.5,3.6
  $\mu$m); (b-f) integrated intensity maps of selected species.
  All panels show the same field of view; (b)\&(c) and (e)\&(f) share colorbars.
  Overlaid are (a) blueshifted/redshifted $^{12}$CO(3-2) and masers (Class II
  ($\Diamond$) and Class I ($+$) CH$_{3}$OH, \citealt{maserpap}; H$_{2}$O ($\bigtriangleup$), \citealt{HC96,Breen11}) and (f) VLA 1.3cm continuum contours.  (g-h) Observed MM2 SED, overplotted with graybody
  fits from our $\beta$-temperature grid (\S\ref{mm2_prop}): in the
  $\chi^2$ surface plot (i), the area below the black line is excluded
  by our 4$\sigma$ 1.3 cm limit.  Levels: 1.3mm: [5,25]$\times\sigma$,
  $\sigma=$0.7 mJy beam$^{-1}$; 1.3cm: [5,8]$\times\sigma$,
  $\sigma=$75 $\mu$Jy beam$^{-1}$; $^{12}$CO: 0.8 Jy beam$^{-1}$ km
  s$^{-1}\times$[5,10,15](blue), $\times$[5,10,15,20,25](red).}
\label{multipanel_fig}
\end{figure}

\begin{figure}
\centerline{\includegraphics[scale=0.7, angle=90]{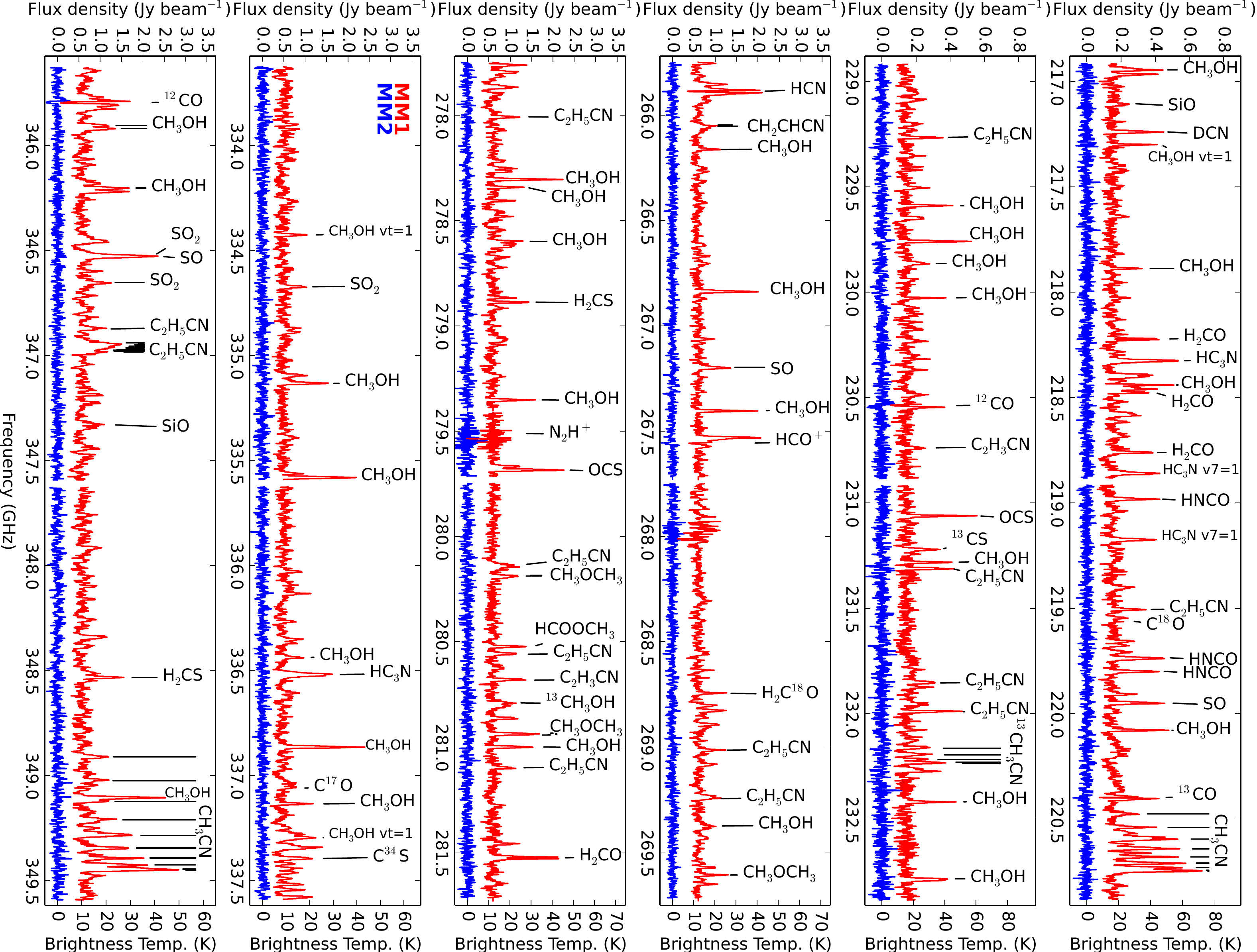}}
\caption{\scriptsize {Continuum-subtracted SMA spectra towards the MM1 (red) and MM2 (blue) continuum peaks, showing the full 24 GHz SMA bandwidth.  MM1 spectra are offset for clarity.  Transition frequencies for selected molecules are labeled for reference.}}
\label{cont_peak_spectra_fig}
\end{figure}

The SMA spectral
setups (and so the nondetections towards MM2) include tracers of cold,
dense gas as well as of rich hot-core chemistry.  In particular, the
1.1 mm tuning was chosen to cover three diagnostic lines:
(1) N$_{2}$H$^{+}$(3-2) (E$_{\rm upper}\sim$ 27 K, n$_{\rm
  crit}\sim$3$\times$10$^{6}$ cm$^{-3}$), for a cold ($<$20 K) core with CO freezeout; (2)
HCO$^{+}$(3-2) (E$_{\rm upper}\sim$ 26 K, n$_{\rm
  crit}\sim$4$\times$10$^{6}$ cm$^{-3}$), for a warmer core
in which CO has come off the grains; and (3) HCN(3-2) (E$_{\rm
  upper}\sim$ 26 K, n$_{\rm crit}\sim$8$\times$10$^{7}$ cm$^{-3}$), for
a very high-density core.  All of these lines are undetected towards MM2, as are (at 1.3 mm) N$_{2}$D$^{+}$(3-2), a tracer of
the inner ``deuteration zone'' in low-mass starless cores
\citep{Caselli02,ppvi_deut}, $^{12}$CO(2-1) and its
isotopologues (Fig.~\ref{multipanel_fig},\ref{cont_peak_spectra_fig}).

Our VLA NH$_{3}$ data provide an independent line of evidence, and
sensitivity to emission from lower-density gas and on larger spatial
scales (Table~\ref{obs_table}).  Like N$_{2}$H$^{+}$, NH$_{3}$ does
not deplete onto grains at densities n$_{\rm H_2}\lesssim$10$^{6}$
cm$^{-3}$ \citep{Bergin97}; however, n$_{\rm crit}$ for the 1.3 cm
NH$_{3}$ inversion transitions is $>$2 orders of magnitude lower than
for N$_2$H$^{+}$(3-2).  NH$_{3}$ emission is detected in the vicinity of MM2, but does not
appear to peak on the millimeter core (Fig.~\ref{multipanel_fig}, showing v=30-39 km s$^{-1}$; see also \S\ref{chem_models}).
In contrast, MM1 is associated with a compact NH$_{3}$ core.

In addition to exhibiting much richer chemistry, MM1 differs from
MM2 in driving a molecular outflow.  As shown in
Figure~\ref{multipanel_fig}, MM1 drives a well-collimated bipolar
molecular outflow, traced by high-velocity $^{12}$CO(3-2) emission.
SiO(8-7) and SiO(5-4) are also detected towards MM1, indicative of recently shocked gas, and
hence an \emph{active} outflow \citep{PdF97,C12}.  The MM1 outflow was previously imaged, at
lower resolution, in $^{12}$CO(2-1), HCO$^{+}$(1-0), and SiO(2-1) by
\citet{C11}.  The new, higher-resolution data better resolve the
outflow lobes, and clearly show that there is no evidence for an outflow driven by MM2.

\section{Discussion: The Nature of MM2}\label{discussion}

\subsection{An Extragalactic Interloper?}\label{dis_xgal}

In the absence of (sub)millimeter line emission, we must consider the chance
of MM2 being an extragalactic background source.  The \citet{Mocanu13} catalog, covering 771 deg$^{2}$ of the South Pole Telescope
Sunyaev-Zel'dovich (SPT-SZ) survey, provides an excellent reference
point.  The SPT-SZ survey is multi-wavelength
($\lambda=$3.2, 2.0, 1.4 mm), allowing dusty sources to be distinguished
from those dominated by synchrotron emission.  The (sub)millimeter-wavelength
spectral index and centimeter-wavelength nondetection of MM2
are inconsistent with synchrotron emission, so
SPT-SZ dust-dominated sources are the relevant comparison.  At
2\farcs4-resolution, the 1.3 mm flux density of MM2 is 0.203$\pm$0.014
Jy \citep{C11}.  
Considering all dust-dominated SPT-SZ sources, the number density
for S$_{\rm 1.4mm}>$0.2 Jy is $\sim$0.002 deg$^{-2}$
\citep[][Table 9]{Mocanu13}, corresponding to
  4$\times$10$^{-7}$ sources expected within the 1.3 mm SMA primary
  beam (FWHP).  It is thus extremely unlikely that MM2 is a background
  extragalactic source, and we conclude that MM2 is a member of the
  G11.92$-$0.61 (proto)cluster.

\subsection{Physical Properties: Estimates from Dust Emission}\label{mm2_prop}

The combination of physical properties inferred for MM2 from its (sub)millimeter-wavelength
continuum emission is extraordinary.  We estimate the core gas mass
from the observed integrated flux densities using a simple model of
isothermal dust emission, correcting for the dust opacity
\citep[][Equation 3]{C11}.  These estimates, for a range of adopted
dust temperatures, are presented in Table~\ref{cont_table}.  Estimates
for MM1 are included for comparison: for the hot core, the minimum and
maximum adopted temperatures correspond to the two components required
to fit the J=12-11 CH$_{3}$CN spectrum \citep[using the method
  of][]{Hunter14}.

A strict lower limit to the physical temperature of MM2 is provided by
its continuum brightness temperature in the 1.3 mm VEX image: 10.8 K.
To constrain the dust temperature and opacity index ($\beta$), we fit
the three (sub)millimeter flux densities (measured from the convolved
images, \S\ref{results_cont}) and their uncertainties with a
single-temperature modified graybody for each point in a
$\beta$-temperature grid ($\beta=$0.5-3.05, $\Delta\beta=$0.05;
T$=$12-35 K, $\Delta$T=1 K).  For each $\beta$-temperature
combination, the only free parameter is $\tau_{\rm 1.3mm}$: the source
size is fixed to 0\farcs58, the geometric mean of the fitted size at
1.1 mm.  Example graybody fits are shown in
Figure~\ref{multipanel_fig}, along with the $\chi^2$ surface plot for
the $\beta$-temperature grid.  Notably, our 1.3 cm upper limit
independently excludes low-$\beta$ models
(Fig.~\ref{multipanel_fig})\footnote{Upper limits are plotted, but not
  used in the graybody fits: 4$\sigma$ VLA limits at 0.7, 3.6, and
  1.3 cm \citep[][this work]{maserpap,C11vla}, MIPSGAL 24 $\mu$m limit
  (maximum flux density of a point source that would produce the
  observed intensity of the pixel coincident with MM2).}, suggesting
that MM2's moderate (sub)millimeter spectral index
($\alpha=$2.6$\pm$0.1) is due primarily to high optical depth, as
opposed to e.g.\ large grains \citep{Tobin13}.

As illustrated in Figure~\ref{multipanel_fig}(i), the models that best
fit MM2 span a fairly narrow temperature range, $\sim$17-19 K.
Adopting T$_{\rm dust}=$20 K as an approximate upper limit (see also
\S\ref{chem_models}), M$_{\rm gas}\gtrsim$30 M$_{\odot}$ (estimated
from the 1.3 mm VEX data; Table~\ref{cont_table}).  Importantly, this
mass is condensed into a
radius\footnote{\begin{math}r\equiv\frac{1}{2}\sqrt{\theta_{\rm
        min}\theta_{\rm maj}}\end{math}} $<$1000 AU.  Assuming
spherical symmetry, this implies N$_{\rm H_2}>$10$^{25}$ cm$^{-2}$ and
n$_{\rm H_2}>$10$^{9}$ cm$^{-3}$.  The luminosity estimate from the
SED yields L/M$\sim$0.1-0.3 (T=17-20 K), indicative of the earliest
phases in models of massive young stellar object evolution
\citep{Molinari08}.

\subsection{Astrochemical Modeling}\label{chem_models}

\begin{figure}
\centerline{\includegraphics[scale=0.46]{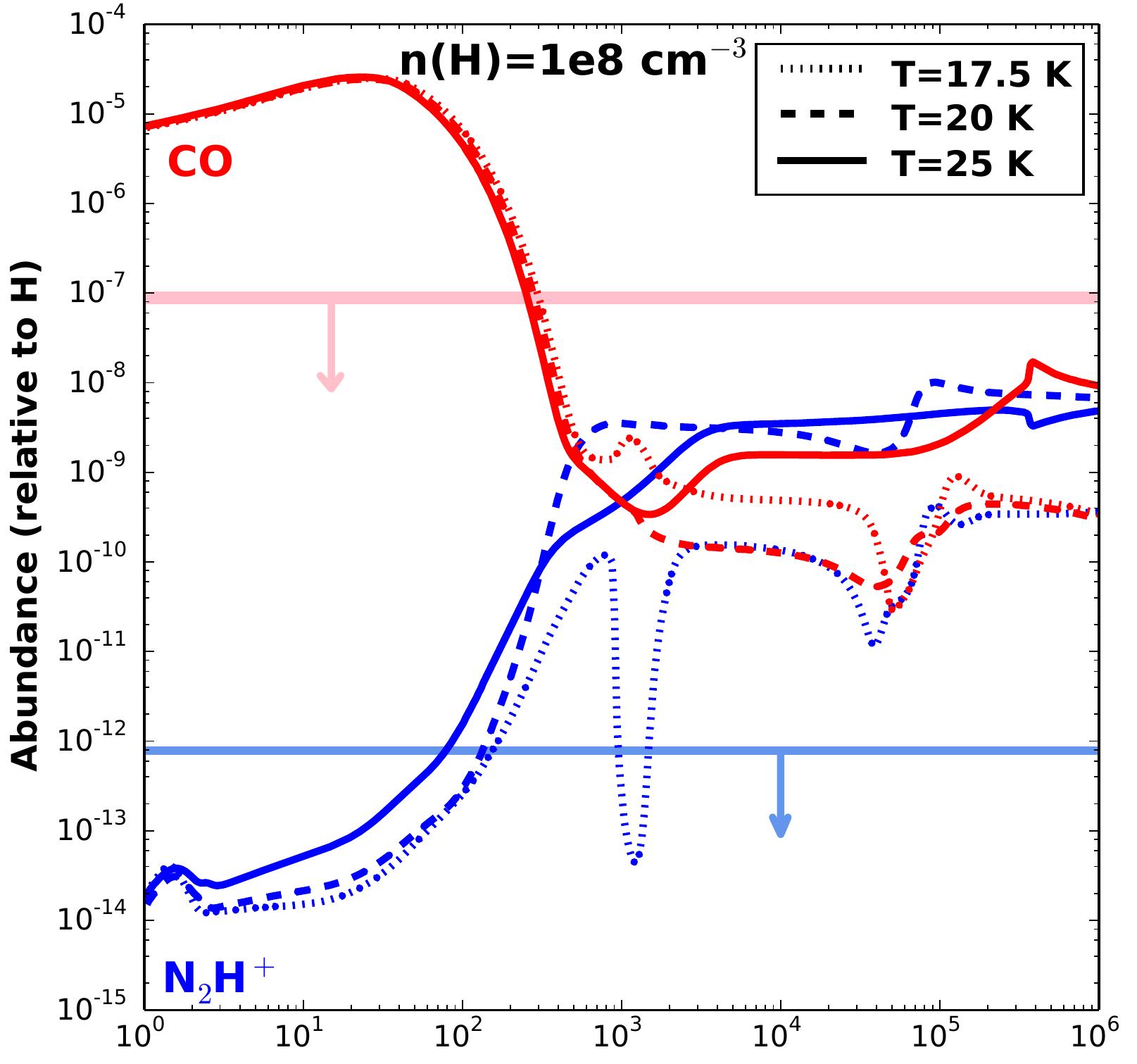}\includegraphics[scale=0.46]{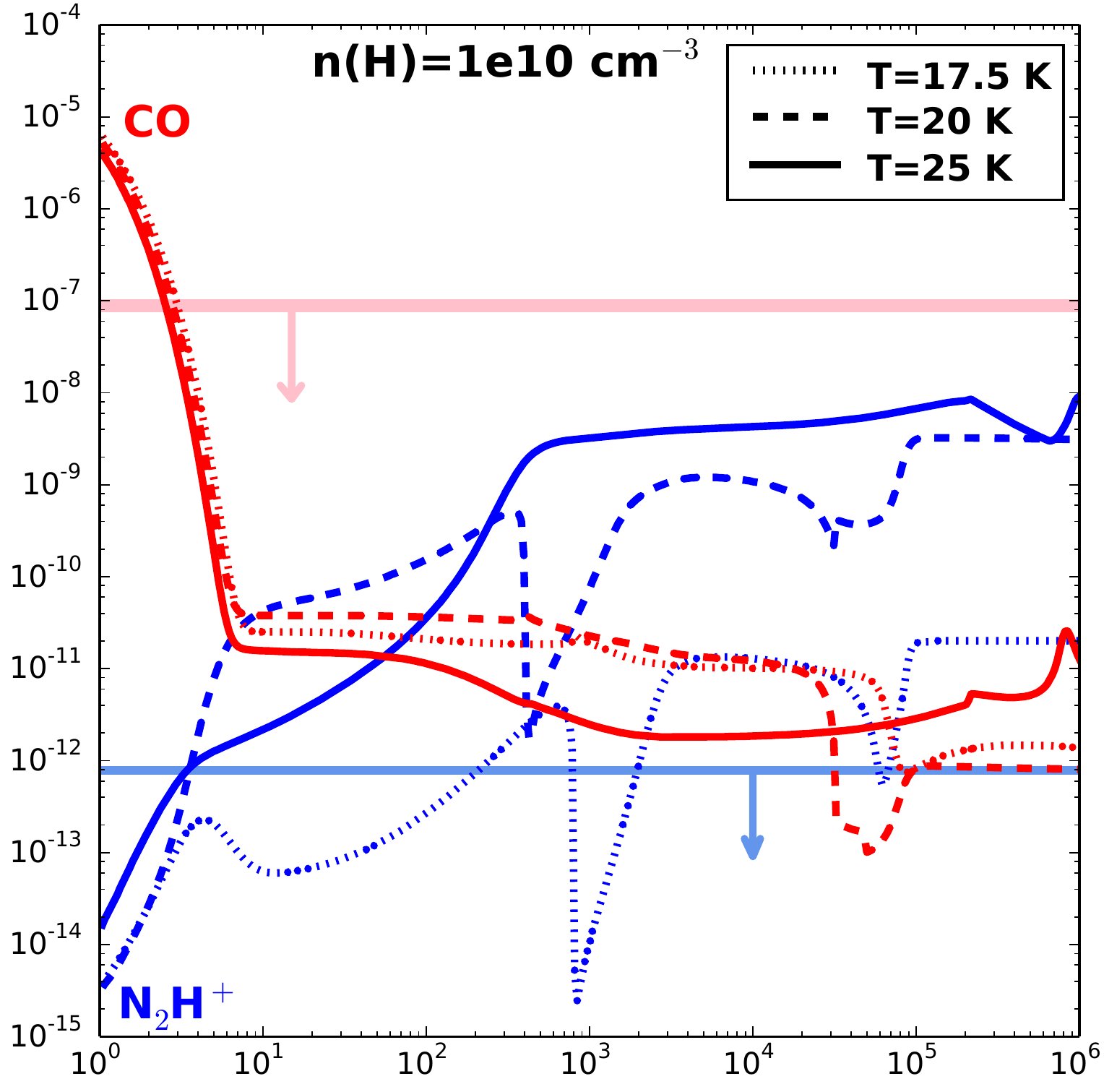}}
\centerline{\includegraphics[scale=0.46]{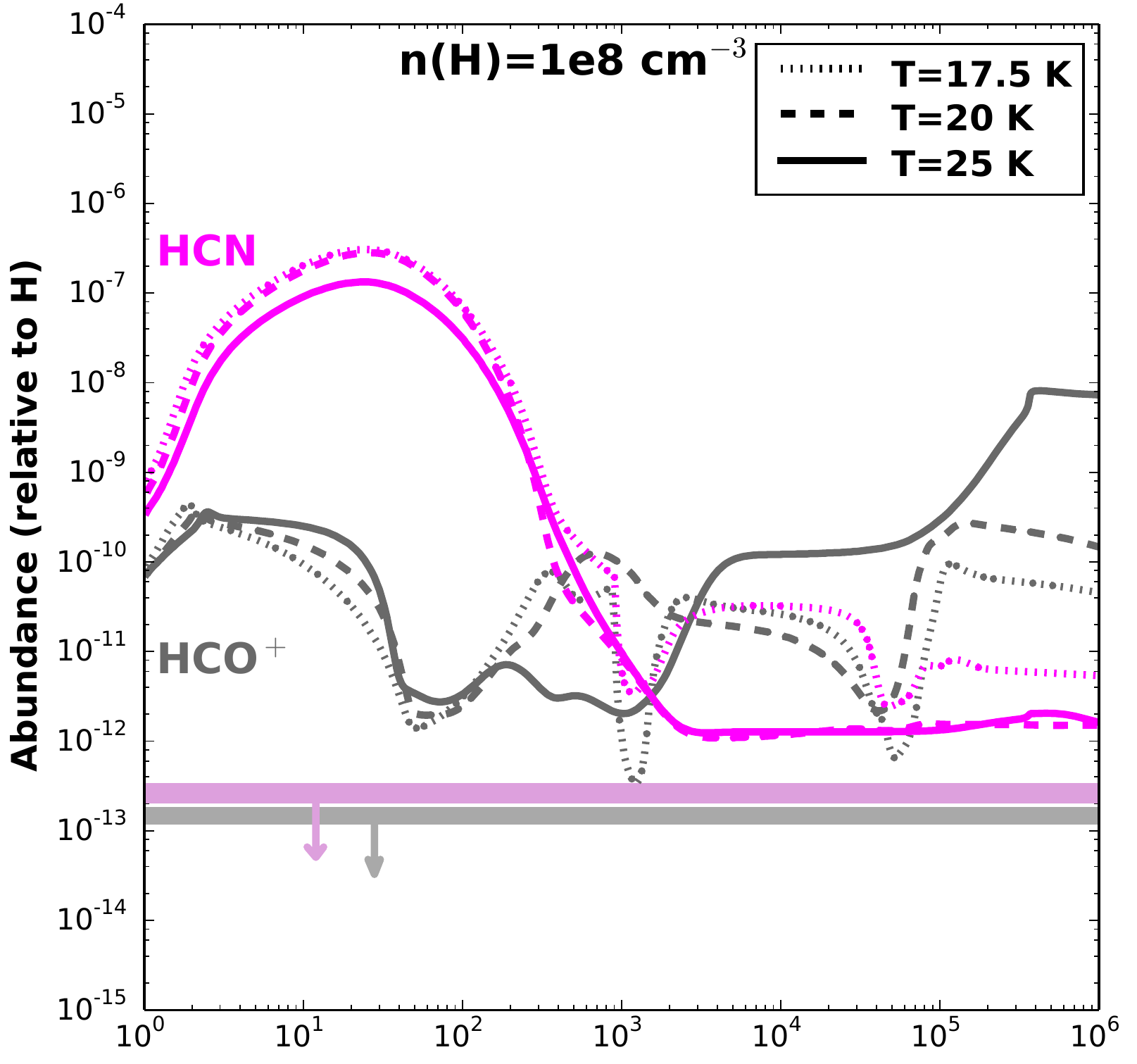}\includegraphics[scale=0.46]{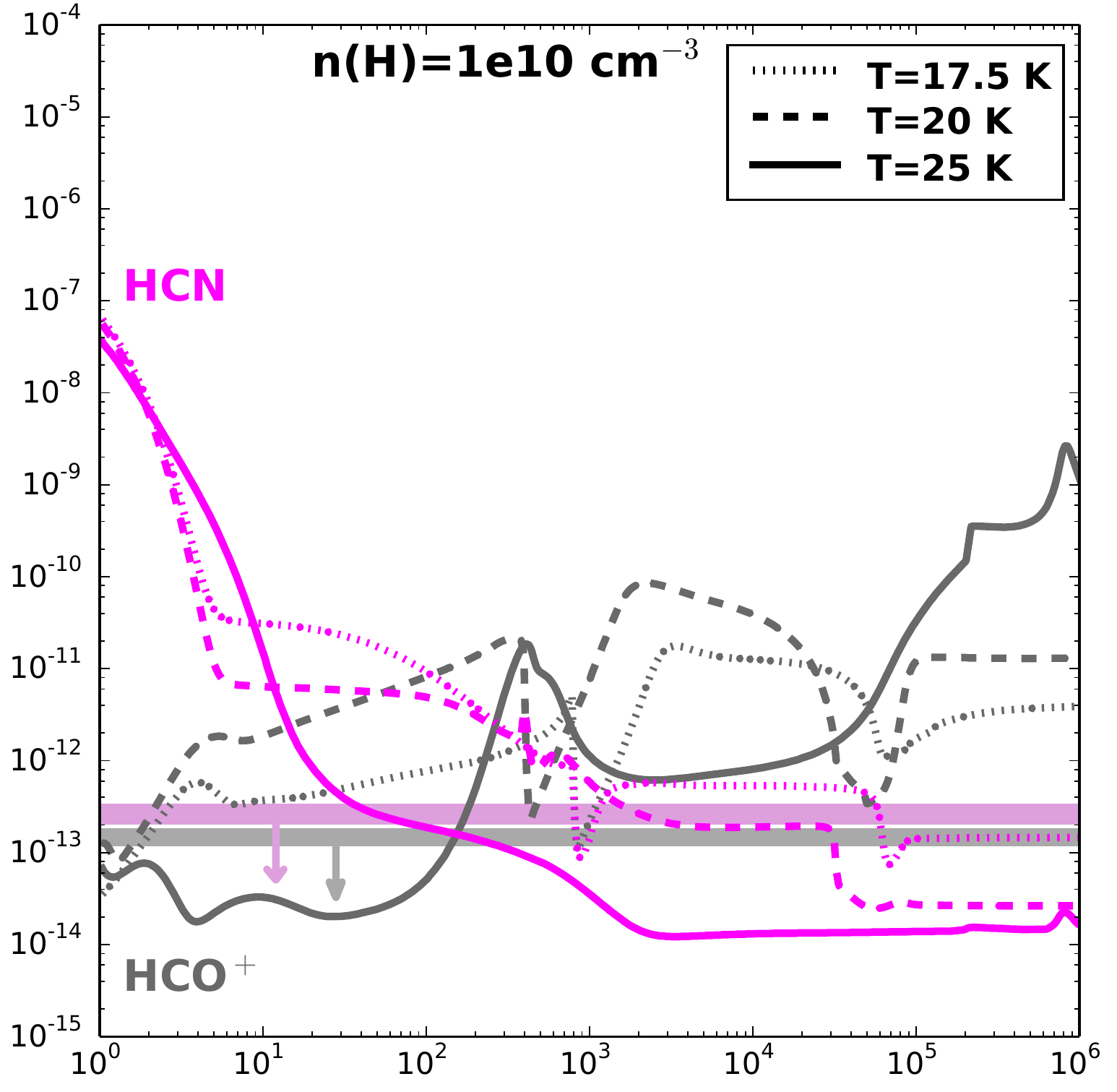}}
\centerline{\includegraphics[scale=0.46]{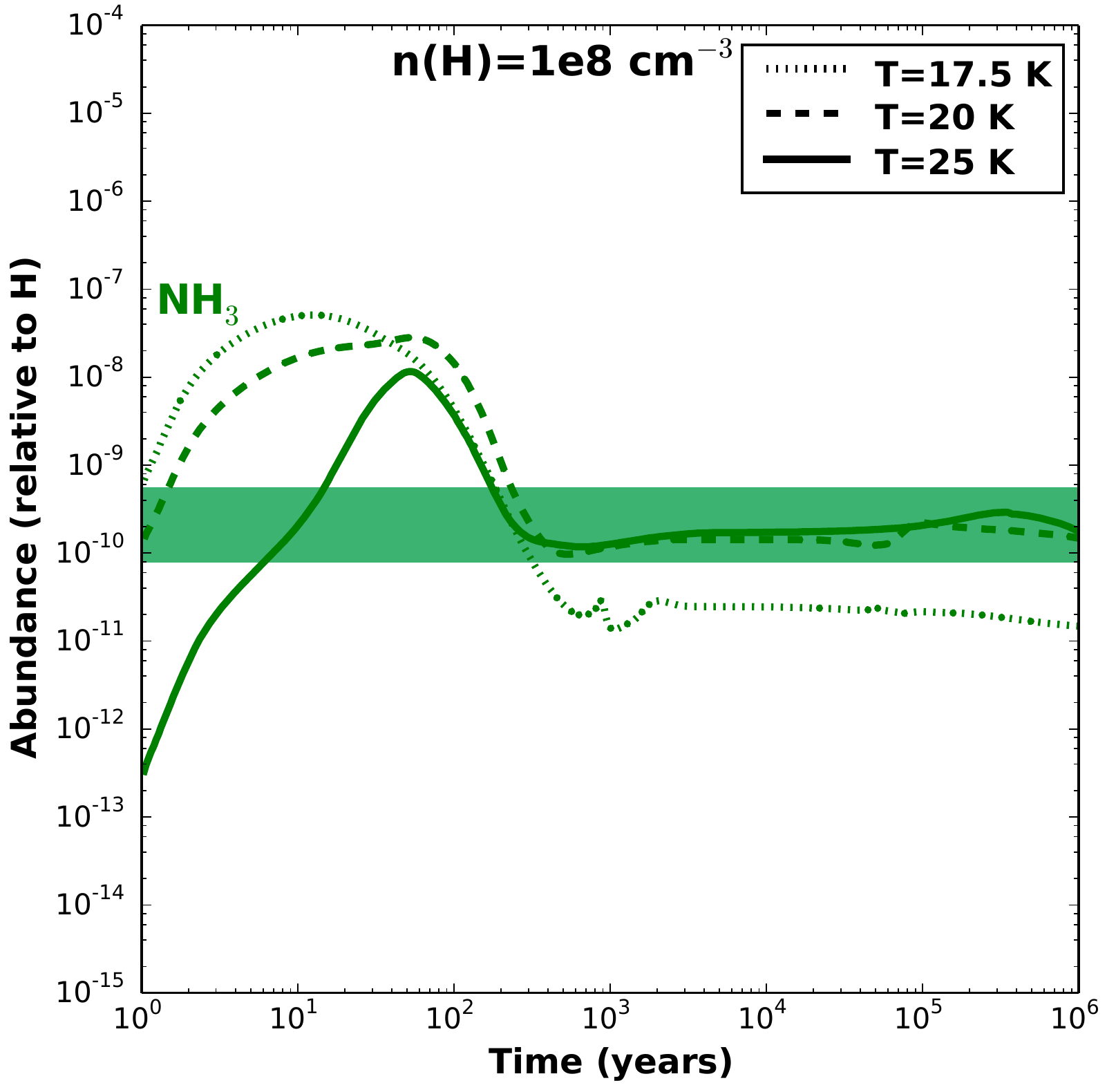}\includegraphics[scale=0.46]{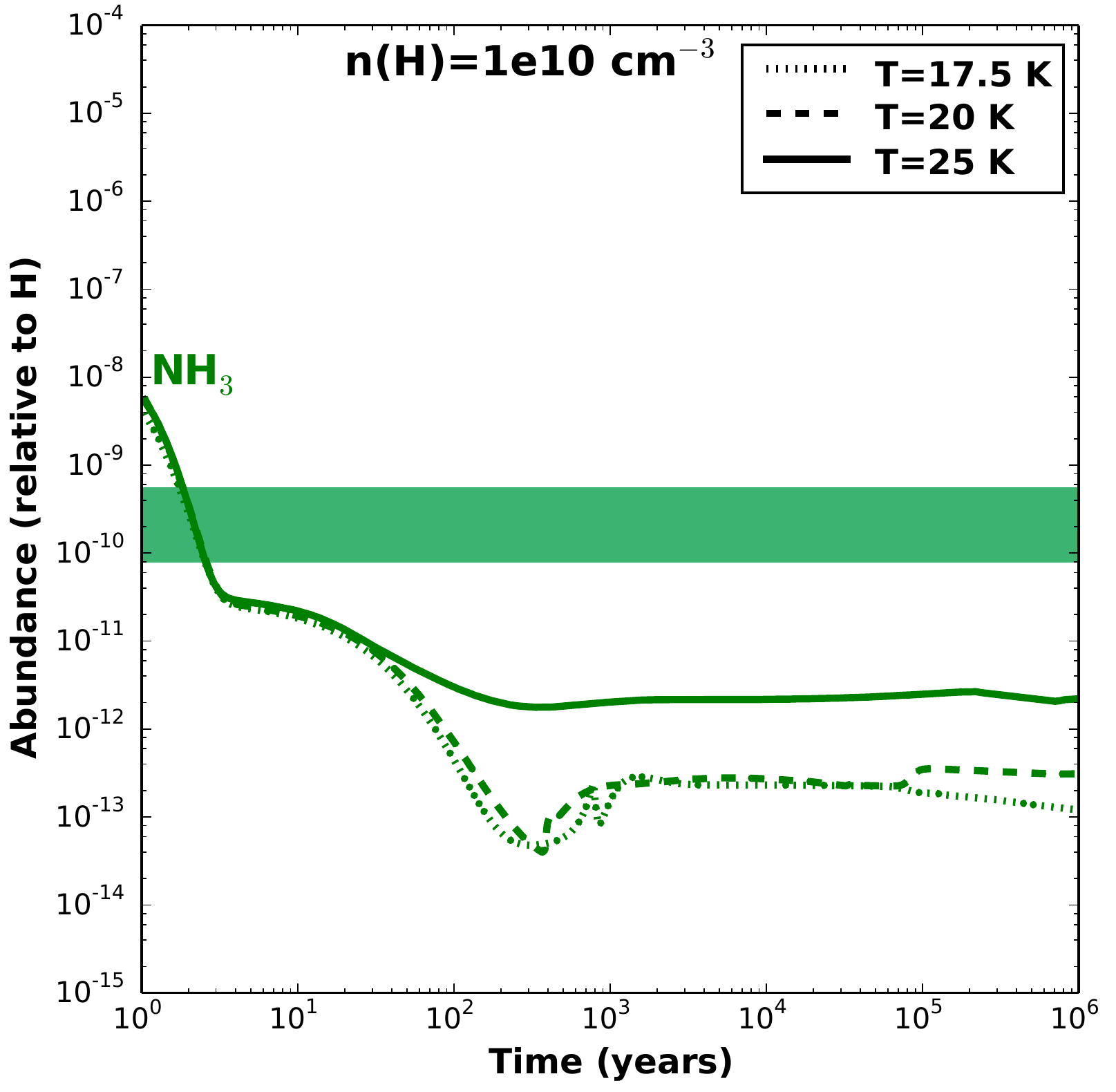}}
\caption{\scriptsize Time-dependent molecular abundances from gas-grain chemical models (A$_{\rm V}=$100); horizontal bands show observed abundances and upper limits (with arrows). Model predictions and observations should only be compared at t$\gtrsim$10$^3$ years, since the meaning of model predictions at shorter times is unclear. Observed N$_2$H$^+$ and HCN limits are only reproduced at low temperatures and high densities, respectively.
}
\label{astrochem_fig}
\end{figure}

The dearth of line emission towards MM2 suggests extreme depletion of gas-species due to
freeze-out onto grains, as seen in cold, dense 
low-mass prestellar cores \citep[n$_{\rm H,max}\sim$10$^{8}$ cm$^{-3}$,][]{Caselli11,ppvi_deut}.  To explore the possible physical properties consistent with the lack of molecular emission lines, we ran a grid of gas-grain
astrochemical models using MONACO \citep{Vasyunin09} and the
Ohio State University gas-grain reaction network \citep{Garrod08}. We used MONACO to model the time evolution of gas-phase and grain surface chemistry under a fixed set of physical conditions. All models were run with oxygen-rich,
low-metal elemental abundances \citep{Graedel82}, a standard dust-to-gas mass ratio and surface site
density \citep[0.01 and 1.5$\times$10$^{15}$ sites cm$^{-2}$,
respectively;][]{Semenov03}, a 10\% reactive desorption efficiency
\citep{Vasyunin13} and a standard molecular cloud cosmic ray ionization (CRI) rate
\citep[$\zeta_{\rm CRI}=$1.3$\times$10$^{-17}$ s$^{-1}$, attenuated from the diffuse ISM;][]{Vasyunin09,Padovani09}.
The standard $\zeta_{\rm CRI}$ is probably too high for this source, since at the very high column densities inferred from the dust emission additional CR attenuation is expected, as discussed below. The model grid comprises seven temperatures (10-25 K, $\Delta$T=2.5 K),
five densities\footnote{n(H)$_{\rm total}=$n(H)+2$\times$n(H$_{2}$)}
(1$\times$10$^{8}$-1$\times$10$^{10}$ cm$^{-3}$), and four A$_{\rm V}$
(10, 100, 1000, 10000).  Increasing A$_{\rm V}$ above 100 has no
effect; Figure~\ref{astrochem_fig} presents results for A$_{\rm V}=$100.  

To quantitatively compare MM2's chemistry with the gas-grain models,
we estimate molecular abundance limits from the SMA data for four
potentially diagnostic species within our observing
bands.\footnote{$\chi$(CO) is estimated from C$^{18}$O(2-1) assuming
  N(CO)/N(C$^{18}$O)=336 \citep{Wilson94}.}  Using RADEX
\citep{radex}, we find the molecular column density equivalent to our
4$\sigma$ T$_{\rm B}$ limit for an assumed linewidth of 2 km s$^{-1}$
and a suite of temperatures (15-25 K, $\Delta$T=2.5 K) informed by the
SED modelling.  For N$_{2}$H$^{+}$, this yields column density limits
only for T$\ge$20 K; if T$<$20 K the SMA surface brightness
sensitivity is insufficient to detect even optically thick
emission.\footnote{N$_{2}$H$^{+}$ is undetected towards MM1, as
  expected for a hot core: once CO desorbs, the
  N$_{2}$H$^{+}$+CO$\rightarrow$HCO$^{+}$+N$_{2}$ destruction channel
  becomes active.}
Molecular abundances, $\chi$(MOL)\begin{math}=\frac{\rm N(MOL)}{\rm N(H)} \end{math}, are calculated adopting N(H)=2$\times$N(H$_{2}$), with N(H$_{2}$) calculated from the
(sub)millimeter continuum emission in the same SMA dataset as the line
of interest (e.g.\ 1.1 mm for N$_{2}$H$^{+}$) for each (assumed)
temperature.  Figure~\ref{astrochem_fig} presents the resulting limits as horizontal stripes: vertical extent reflects the range in assumed temperature.
We emphasize that there are substantial inherent
uncertainties (e.g.\ linewidth, temperature, estimation of
N(H$_{2}$)), and the derived abundance limits should be considered
order-of-magnitude estimates.  For NH$_{3}$, additional uncertainty is
introduced by the difference in morphology between the NH$_{3}$ and
dust continuum emission (Fig.~\ref{multipanel_fig}) and the difference
in \emph{uv}-coverage between the VLA and SMA data.  We use the VLA spectra at the MM2 continuum peak
to estimate N(NH$_{3}$) \citep[following][]{C13}, and calculate $\chi$(NH$_{3}$)
for T$_{\rm dust}=$15-25 K using two different measures of N(H$_{2}$):
our 1.1 mm SMA data and the 1.3 mm compact-configuration SMA data from
\citet{C11}.  Figure~\ref{astrochem_fig} shows the resulting $\chi$(NH$_{3}$) range (green rectangles).
   
Comparing the data and models after the initial freezeout
(Fig.~\ref{astrochem_fig}; t$\gtrsim$10$^3$ years) shows that it is possible to explain the observed dearth of molecular emission assuming a standard gas-grain chemistry if the MM2 conditions are indeed as extreme as suggested by the SED modeling. Furthermore, specific model predictions provide three key
insights.  
(1) The non-detection of N$_{2}$H$^{+}$ rules out T$\ge$20
K for MM2.  For T$\ge$20 K, the model-predicted N$_{2}$H$^{+}$
abundances exceed the observational limit by $\gtrsim$3 orders of
magnitude.
(2) The non-detection of HCN favors the highest density models.  For
T$<$20 K and n(H)$\sim$10$^{10}$ cm$^{-3}$, the
model-predicted HCN abundances are consistent with the observational
limit, within the observational uncertainties and 
$\sim$order-of-magnitude uncertainty intrinsic in the model results
\citep{Vasyunin04,Vasyunin08}.  Lower density models ($\sim$10$^{8}$
cm$^{-3}$) predict higher $\chi$(HCN), above the observed limit for all
times and temperatures.
(3) The NH$_{3}$ detected with the VLA near MM2 cannot
trace the dense gas seen in dust continuum emission with the SMA.  The
inferred NH$_{3}$ abundance is $\sim$10$^{3}\times$ higher than
predicted by the cold, dense models required by the SMA nondetections
of N$_{2}$H$^{+}$ and HCN; instead, $\chi$(NH$_{3}$) is
consistent with comparatively lower-density ($\sim$10$^{8}$ cm$^{-3}$),
warmer (20-25 K) material.  From simple estimates, external
heating by MM1/MM3 \citep[L$\sim$10$^{4}$ L$_{\odot}$,][]{C11} is sufficient to
account for NH$_{3}$ temperatures of $\sim$20 K near MM2. 

The poor model-data agreement for HCO$^+$ even at high densities
suggests that $\zeta_{\rm
  CRI}$ is too high (the agreement is worse for lower-density models): a lower CRI rate \citep[e.g.\ attenuated within
  the high-density core,][Fig.~1]{Padovani13} would reduce the
model-predicted abundances.  Overestimated desorption rates could,
however, also overpredict $\chi$(HCO$^+$).  Evaluating the level of CR
attenuation would require new, deep observations of the only
observable ions expected not to freeze out under any temperature and
density conditions: H$_2$D$^+$ and D$_2$H$^+$ \citep[e.g.][]{Ceccarelli05}.

\subsection{The Best Candidate?}

Based on the available evidence, MM2 is the best candidate for a
massive starless core discovered to date.  Crucially, MM2 has a
centrally condensed mass sufficient to form a massive star:
$\gtrsim$30 M$_{\odot}$ within
R$<$1000 AU.  Compared to CygX-N53-MM2 (another contender for
``best candidate''), G11.92$-$0.61--MM2 has $\gtrsim$5$\times$
as much mass on $\sim$1500 AU sizescales \citep[estimated from millimeter continuum emission;][]{Bontemps10}.  \citet{DuarteCabral13}
also report a tentative outflow detection towards CygX-N53-MM2,
raising the question of whether it is truly starless.  Compared to the \citet{Tan13}
cores, G11.92$-$0.61--MM2 is more massive
(with the possible exception of C1-S), more compact, and several
orders of magnitude denser. 
G11.92$-$0.61--MM2 has also been more extensively and sensitively
searched for star formation indicators.

Interestingly, \citet{Kauffmann13} recently proposed that very short
lifetimes could explain the lack of observational examples of massive
starless cores, consistent with the very short free-fall time
($\lesssim$1,000 years) implied by MM2's high density.  Further
comparison of MM2's properties with the predictions of core accretion
models for massive star formation requires additional observations,
e.g. of the dense-core tracers H$_{2}$D$^{+}$, N$_{2}$H$^{+}$, and
N$_{2}$D$^{+}$ with the sensitivity of ALMA.  Most important is
detecting line emission from the dense, cold millimeter core to
determine whether MM2 is gravitationally bound and will collapse to
form a massive star.

\acknowledgments

Supported by NSF AAPF (C.J.C, AST-1003134) and ERC (A.V., PALs 320620). 
C.J.C. thanks E.\ Rosolowsky, K.\ Rowlands, and A.-M.\ Weijmans.

\begin{deluxetable}{lcccc}
\rotate
\tablewidth{0pt}
\tablecaption{Observing Parameters \label{obs_table}}
\tablehead{
\colhead{Parameter} &
\colhead{SMA 1.3 mm} & 
\colhead{SMA 1.1 mm} & 
\colhead{SMA 0.88 mm} & 
\colhead{VLA 1.3 cm} \\ 
}
\tablecolumns{5}  
\tabletypesize{\scriptsize}
\setlength{\tabcolsep}{0.15in}
\startdata 
Observing date (UT) & 2011 Aug 28 &  2013 May 31 & 2011 Aug 19 & 2010 Aug 25, 2011 Jan 30  \\
Project code        & 2011A-S076  & 2013A-S043   & 2011A-S076  & AB1346 \\
Configuration & Very Extended & Extended & Extended & D$+$CnB \\
Number of antennas &     8      &    7     &    8   &  21, 21  \\
$\tau_{\rm 225GHz}$ &     $\sim$0.05       &     $\sim$0.1-0.15       &   $\sim$0.04-0.08   &  n/a  \\
T$_{\rm sys}$ (source transit) &   $\sim$90 K       &    $\sim$200 K        &    $\sim$200 K &   n/a          \\
Phase Center (J2000): & & & \\
~~~~~R.A. & 18$^{\rm h}$13$^{\rm m}$58$^{\rm s}$.10 & 18$^{\rm h}$13$^{\rm m}$58$^{\rm s}$.10 & 18$^{\rm h}$13$^{\rm m}$58$^{\rm s}$.10  &  18$^{\rm h}$13$^{\rm m}$58$^{\rm s}$.10 \\
~~~~~Dec.                     & $-$18$^{\circ}$54\arcmin16\farcs7 & $-$18$^{\circ}$54\arcmin16\farcs7 & $-$18$^{\circ}$54\arcmin16\farcs7 & $-$18$^{\circ}$54\arcmin17\farcs0 \\
Primary beam size (FWHP) & 52\arcsec & 43\arcsec & 34\arcsec & 1.9\arcmin \\
Frequency coverage: & & &                                                         & mean continuum: 24.85 GHz \\
~~~~~LSB & $\sim$216.9-220.9 GHz & $\sim$ 265.7-269.7 GHz & $\sim$333.6-337.6 GHz & 8$\times$8 MHz \\
~~~~~USB & $\sim$228.9-232.9 GHz & $\sim$ 277.7-281.7 GHz & $\sim$345.6-349.6 GHz & 8$\times$8 MHz \\
Channel width\tablenotemark{a} & 0.8125 MHz & 0.8125 MHz & 0.8125 MHz & 31.25 kHz \\
                               &            & 0.406 MHz (N$_{2}$H$^{+}$) &   &   (0.4 km s$^{-1}$)        \\
Resampled velocity resolution &   1.12 km s$^{-1}$       &     0.92 km s$^{-1}$                  &    0.75  km s$^{-1}$        &   n/a     \\
Gain calibrators & J1733-130, J1924-292 & J1733-130, J1924-292 & J1733-130, J1924-292& J1820-2528 \\
Bandpass calibrator & 3C84 & 3C279 & 3C84 & J1924-292\\
Flux calibrator & Callisto\tablenotemark{b} & Neptune\tablenotemark{b} & Callisto\tablenotemark{b} & 3C286 \\
Angular resolution\tablenotemark{c} & 0\farcs574$\times$0\farcs370 (P.A.=30$^{\circ}$)& 0\farcs906$\times$0\farcs837 (P.A.=$-$74$^{\circ}$) & 0\farcs798$\times$0\farcs703 (P.A.=54$^{\circ}$) & 1\farcs077$\times$0\farcs833 (P.A.=$-$35$^{\circ}$)\\
Largest angular scale\tablenotemark{d} & 9\arcsec & 9\arcsec & 9\arcsec & $\sim$40\arcsec \\
Projected baselines                    &  $\sim$20-394 k$\lambda$        &   $\sim$17-191 k$\lambda$        &     $\sim$18-225 k$\lambda$        &  $\sim$1.8-315 k$\lambda$          \\          
Continuum rms noise\tablenotemark{c} & 0.7 mJy beam$^{-1}$ & 3 mJy beam$^{-1}$  & 3 mJy beam$^{-1}$ & 75 $\mu$Jy beam$^{-1}$ \\
Spectral line rms noise\tablenotemark{e}& 23 mJy beam$^{-1}$ & 71  mJy beam$^{-1}$ & 73  mJy beam$^{-1}$ & 1.5 mJy beam$^{-1}$ \\
                                        &                    & 150 mJy beam$^{-1}$  (N$_{2}$H$^{+}$)\tablenotemark{f}       &  55 mJy beam$^{-1}$ ($^{12}$CO) & \\ 
\enddata 
\tablenotetext{a}{1.1mm: of 48 correlator ``chunks'', two have 0.406MHz and six have 1.625MHz channels.} 
\tablenotetext{b}{Using Butler-JPL-Horizons 2012 models.}
\tablenotetext{c}{Combined LSB+USB continuum image (Briggs weighting, robust$=$0.5).  VLA NH$_3$ images were made with robust$=$1.0 for best sensitivity.}
\tablenotetext{d}{Scale at which 10\% of peak brightness would be recovered for a Gaussian source; for 50\% recovery, multiply by 0.55 \citep{WW94}.  VLA: Estimate for combined D+CnB images, which are dominated by the higher-weight CnB data.}
\tablenotetext{e}{Typical rms per channel; hanning-smoothed.}
\tablenotetext{f}{Near V$_{\rm LSR}\sim$35 km s$^{-1}$.}

\end{deluxetable}

\begin{deluxetable}{lccccccccccccc}
\rotate
\tablewidth{0pt}
\tablecaption{Properties of Continuum Sources \label{cont_table}}
\tablehead{
\colhead{} & 
\multicolumn{5}{c}{Observed Properties} &
\colhead{} &
\colhead{} &
\colhead{} &
\colhead{} &
\multicolumn{4}{c}{Derived Properties}\\
\colhead{Source} &
\multicolumn{2}{c}{J2000 Coordinates\tablenotemark{a}} &
\colhead{Peak Intensity\tablenotemark{a}} & 
\colhead{Integ. Flux\tablenotemark{a}} &
\colhead{Size\tablenotemark{a}} & 
\colhead{Size} & 
\colhead{T$_{b}$} &
\colhead{T$_{\rm dust}$} &
\colhead{$\tau_{\rm dust}$} & 
\colhead{$\kappa_{\nu}$\tablenotemark{b}} &
\colhead{M$_{\rm gas}$} & 
\colhead{N$_{H_{2}}$} & 
\colhead{n$_{H_{2}}$} \\ 
\colhead{} & 
\colhead{$\alpha$ ($^{\rm h}~~^{\rm m}~~^{\rm s}$)} &
\colhead{$\delta$ ($^{\circ}~~{\arcmin}~~{\arcsec}$)} &
\colhead{(mJy~~~~~~~~~} & 
\colhead{Density (mJy)} & 
\colhead{(\arcsec $\times$ \arcsec [P.A.$^{\circ}$])} &
\colhead{(AU $\times$ AU)} & 
\colhead{(K)} &
\colhead{(K)} &
\colhead{} &
\colhead{(cm$^{2}$ g$^{-1}$)} & 
\colhead{(M$_{\odot}$)} &
\colhead{$\times$10$^{25}$} &
\colhead{$\times$10$^{9}$}  \\
\colhead{} & 
\colhead{} & 
\colhead{} & 
\colhead{beam$^{-1}$)} & 
\colhead{} & 
\colhead{} & 
\colhead{} & 
\colhead{} & 
\colhead{} & 
\colhead{} & 
\colhead{} & 
\colhead{} & 
\colhead{(cm$^{-2}$)} & 
\colhead{(cm$^{-3}$)}  
}
\tablecolumns{14}  
\tabletypesize{\scriptsize}
\setlength{\tabcolsep}{0.05in}
\startdata 
\sidehead{\bf SMA 1.3 mm VEX}
MM1 & 18 13 58.1099 & -18 54 20.141 & 90 (1) &  138 (2) & 0.34 (0.01) $\times$ 0.29 (0.01) [125 (7)] & 1150 $\times$ 960 & 34.0 & 150-240 & 0.3-0.2  & 1.11 & 3-2  & 0.9-0.5  & 0.9-0.5 \\ 
\\
MM2 & 18 13 57.8599 & -18 54 13.958 & 44 (1) & 90 (2) & 0.56 (0.02) $\times$ 0.36 (0.03) [71 (5)] & 1880 $\times$ 1220 & 10.8  & 16 & 1.1 & 1.00 & 47  & 6.0  & 4.0  \\
  &                &              &         &       &                                            &                     &       & 17 & 1.0 & 1.00 & 41  & 5.3  & 3.5 \\
  &                &              &         &       &                                            &                     &       & 19 & 0.8 & 1.00 & 33  & 4.3  & 2.8 \\
  &                &              &         &       &                                            &                     &       & 20 & 0.8 & 1.00 & 30  & 3.9  & 2.6 \\

\sidehead{\bf SMA 1.1 mm EX}
MM1 & 18 13 58.1102 & -18 54 20.201 & 295 (5) & 397 (6) & 0.66 (0.02) $\times$ 0.36 (0.04) [130 (2)] & 2210 $\times$ 1200  & 27.7  & 150-240 & 0.2-0.1 & 1.5 & 5-3 & 0.5-0.3 & 0.3-0.2 \\
\\
MM2 & 18 13 57.8568 & -18 54 13.99 &  132 (3) & 191 (5) & 0.61 (0.05) $\times$ 0.55 (0.05) [170 (30)] & 2100 $\times$ 1800  & 9.4  & 16  & 0.9  & 1.36 & 48 & 3.8 & 1.9  \\
 &                &              &         &       &                                            &                     &         & 17 & 0.8  & 1.36 & 43 &   3.3 & 1.7 \\
 &                &              &         &       &                                            &                     &         & 19 & 0.7  & 1.36 & 35 &   2.7 & 1.4 \\
 &                &              &         &       &                                            &                     &         & 20 & 0.6  & 1.36 & 32 &   2.5 & 1.3 \\

\sidehead{\bf SMA 0.88 mm EX}
MM1 & 18 13 58.1066 & -18 54 20.158 & 363 (6) & 510 (8) & 0.58 (0.02) $\times$ 0.34 (0.03) [134 (4)] & 1940 $\times$ 1200  & 27.1 & 150-240  & 0.2-0.1 & 2.3 & 3-2 & 0.3-0.2 & 0.2-0.1 \\
\\
MM2 & 18 13 57.8546 & -18 54 13.92 & 167 (5) & 294 (9) & 0.66 (0.05) $\times$ 0.64 (0.05) [4 (46)] & 2200 $\times$ 2200  & 7.3  & 16  & 0.6  & 2.11 & 31  & 1.9 & 0.9\\
  &               &              &          &         &                                          &                      &     & 17  & 0.6  & 2.11 & 27  & 1.7 & 0.8\\
  &               &              &          &         &                                          &                      &     & 19  & 0.5  & 2.11 & 22  & 1.4 & 0.6\\
  &               &              &          &         &                                          &                      &     & 20  & 0.5  & 2.11 & 20  & 1.3 & 0.6\\
\sidehead{\bf VLA 1.3 cm}
CM1  & 18 13 58.107 &  -18 54 20.20 & 0.63(0.08) & \nodata & $<$0.95 & $<$3200 & & &  &  &  & &\\
\sidehead{\bf SMA 1.3 mm VEX: uvrange $<$191 k$\lambda$, convolved to 1.1 mm beam}
MM1 & 18 13 58.1110 & -18 54 20.144 & 121 (2) & 148 (3) &                                          &                    &     &&&&&& \\
MM2 & 18 13 57.8557 &  -18 54 13.97 & 71 (2) & 103 (3)  &                                          &                    &     &&&&&& \\
\sidehead{\bf SMA 0.88 mm EX: convolved to 1.1 mm beam}
MM1 &  18 13 58.1068 & -18 54 20.160 & 388 (6) &  525 (8)   &                                          &                    &     &&&&&& \\
MM2 &  18 13 57.8532 & -18 54 13.91 & 185 (6) &  302 (9)   &                                          &                    &     &&&&&& \\
\enddata 
\tablenotetext{a}{SMA: from two-dimensional Gaussian fitting; ``size'' is deconvolved source size.  Statistical uncertainties are indicated by the number of significant figures or given in parentheses.  VLA: parameters of peak pixel; quoted uncertainties are one pixel (position) and 1$\sigma$ (peak intensity).}
\tablenotetext{b}{\citet{OH94}: grains with thin (MM1) or thick (MM2) ice mantles and coagulation at 10$^{8}$ cm$^{-3}$; linearly interpolated to 1.1 and 0.88 mm.}

\end{deluxetable}

\end{document}